\documentclass{amsart}
\usepackage{graphicx}

\usepackage{amsmath,amssymb}
\usepackage{latexsym}

\newtheorem{theorem}{Theorem}[section]
\newtheorem{proposition}{Proposition}[section]
\newtheorem{lemma}[theorem]{Lemma}
\newtheorem{corollary}[theorem]{Corollary}

\theoremstyle{definition}

\theoremstyle{remark}

\numberwithin{equation}{section}

\newcommand{\grad}{\nabla}

\newcommand{\R}{\mathbb R}

\newcommand{\N}{\mathbb N}
\newcommand{\C}{\mathbb C}
\newcommand{\imm}{\mathrm{i}}              
\newcommand{\e}{\mathrm{e}}                

\newcommand{\ve}[1]{{\mathbf {#1}}}

\newcommand{\Cal}[1]{{\mathcal {#1}}}
\usepackage{color}

\newcommand{\red}[1]{{\color{magenta}#1}}

\begin{document}

\title[Dephasing of Kuramoto oscillators]{Dephasing of Kuramoto oscillators
in kinetic regime towards a fixed asymptotically free state}


\author{D. Benedetto}
\email{benedetto@mat.uniroma1.it}

\author{E. Caglioti}
\email{caglioti@mat.uniroma1.it}

\author{U. Montemagno}
\email{montemagno@mat.uniroma1.it}
\address{Dipartimento di Matematica, {\it Sapienza} Universit\`a di Roma - 
P.zle A. Moro, 5, 00185 Roma, Italy}



\date{\today}


\begin{abstract}

We study the kinetic Kuramoto model for coupled oscillators.
We prove that for any regular asymptotically free state, 
if the interaction is small enough, 
it exists a solution which is asymptotically close
to it. For this class of
solution the order parameter vanishes to zero, showing
a behavior similar to the phenomenon of Landau damping in plasma
physics. We obtain an exponential decay of the order parameter
in the case on analytical regularity of the asymptotic state, 
and a polynomial decay in the case of Sobolev regularity.
\end{abstract}

\maketitle
\noindent
\section{Introduction}

The Kuramoto model is a mean-field model of coupled oscillators,
proposed by Kuramoto to describe synchronization phenomena (see
\cite{kuramoto},\cite{strogatz2000},\cite{bonilla}).  The equations
for the phases of the oscillators are
\begin{equation}\label{kuram}
 \dot{\vartheta}_i(t)=\omega_i-\frac{\mu}{N}
\sum_{j=1}^{N}\sin({\vartheta}_i(t)-{\vartheta}_j(t)),\quad i=1,\dots ,N;
\end{equation}
where the phases $\vartheta_i$ 
can be considered in the one-dimensional torus $\Cal T$, i.e.
defined mod $2\pi$.
The parameters 
$\omega_i$ are the 'natural frequencies' of the  oscillators and
$\mu>0$ is the coupling intensity.
It can be useful to
represent the system \eqref{kuram} in the unitary circle
in the complex plane
by considering $N$ particles with position $\e^{\imm \vartheta_i(t)}$.
The center of mass  is in the point
\begin{equation}\label{deford}
 R_N(t)e^{\imm\varphi_N(t)}=\frac{1}{N}\sum_{j=1}^Ne^{\imm\vartheta_j(t)},
\end{equation}
where $0 \le R_N(t) \le 1$ and $\varphi_N(t)$ is well defined only if
$R_N(t)>0$. Using this definition the system 
\eqref{kuram} can be rewritten in a clearer form:
\begin{equation}\label{kuram1}
 \dot{\vartheta}_i(t)=\omega_i-
\mu R_N(t)\sin(\vartheta_i(t)-\varphi_N(t)),\quad i=1,\dots, N, 
\end{equation}
as follows from easy calculations.
The interaction term here becomes an attraction term 
towards the phase of the center of mass, 
and the intensity of the attraction is driven by $R(t)$,
 which grows when the particles get closer.

Taking the $N\to +\infty$ limit of the $\eqref{kuram}$ we obtain the following equation:
\begin{equation}\label{noneq}
\begin{cases}
\partial_t f(t,\vartheta,\omega)+ \partial_\vartheta(v(t,\vartheta,\omega)f(t,\vartheta,\omega))=0\\
v(t,\vartheta,\omega)=\omega -\mu\int_{\mathcal{T}\times \R} \sin(\vartheta-\vartheta')f(t,\vartheta',\omega')d\vartheta'd\omega'\\
\end{cases},
\end{equation}
where $f(t,\vartheta,\omega)$ is a probability density in $\Cal T \times \R$.
The distribution of the natural frequencies 
is $g(\omega ) = \int_{\Cal T}  f(t,\vartheta, \omega)d\vartheta$,
which is a conserved quantity.

Existence and uniqueness results for this equation are obtained in \cite{lancellotti} where the \eqref{noneq} is rigorously derived by doing the kinetic limit of \eqref{kuram}.

The \eqref{deford}  passes to the limit giving
\begin{align}\label{orderparam}
&R(t)\e^{\imm \varphi(t)}=\int f(t,\vartheta,\omega)\e^{\imm \vartheta}d\vartheta d\omega, 
\end{align}
so that $v$ can be rewritten in a simpler form:
\begin{align}
v(t,\vartheta,\omega)=\omega-\mu R(t)\sin(\vartheta-\varphi(t)).
\end{align}

The asymptotic behavior of the solutions of  \eqref{kuram} and \eqref{noneq} 
is well understood in the case in which all the oscillators have the same 
natural frequency, i.e. $g(\omega)= \delta (\omega -\bar \omega)$:
the oscillators synchronize in phase (see \cite{dong}, \cite{kuramoto2014}).

Otherwise, the behavior of solutions depends on the value of 
the coupling parameter $\mu$; in particular, 
if $g$ has compact support and $\mu$ is sufficiently large, 
the oscillators synchronize in frequency 
(see \cite{MR2897541}, \cite{carrillo}, \cite{HHK}, 
see also \cite{kuramoto2014}).

The total synchronization is impossible if $g$ has not compact support, 
and in this case it is expected a partial synchronization 
(see \cite{strogatz2000}). Moreover, in \cite{ldstrogatz}
it is firstly proved a Landau-damping type results 
for the kinetic equation \eqref{noneq}, in the linearized 
case: it is shown that,
for a perturbation of a constant density, $R$ decays
sub-exponentially to zero. More recently, 
in \cite{mirolloor} it is shown that 
an exponential decay of $R$ can only be
obtained if $g(\omega)$ is analytic (while in \cite{ldstrogatz}
the support of $g$ is compact).

A full Landau-damping type results for the non linear case
have been obtained in the recent works \cite{giacomin} 
(in the case of Sobolev regularity) and 
\cite{dephforw} (in the case of analytical 
regularity): 
for regular initial data and sufficiently small coupling 
parameter, it is
shown that the order parameter $R$ vanishes to zero, and the solution
$f$ is asymptotically close to a free flow. 

In this work we show a complementary result: we fix an  asymptotically free
flow and we find a solution which converges to it. We 
take inspiration from the paper  \cite{cm}, where 
it is proved a similar Landau damping type result 
for the Vlasov-Poisson equation 
(for a reference about the Landau damping for Vlasov type equation 
see \cite{villani},\cite{ldexp},\cite{ldsobolev}).
As in \cite{cm}, we obtain exponentially fast convergence
in the case of analytical regularity of the asymptotic state, 
but here we can also consider the case of Sobolev regularity 
of the asymptotic state, which is the main novelty in this work.

\section{Exponential dephasing}\label{sez:exp}

In this section we prove our first result for the equation
\eqref{noneq} where the asymptotic datum $f_\infty$ is analytic and
the decaying of $R$ is exponential.  The precise statement of the
result is:

\begin{theorem}\label{backwardld}
Let $f_\infty: \Cal T\times \R\longrightarrow \R^+$ be a
probability 
density which satisfies \\
$\sup_{k,\eta}\left|{\hat f}_{\infty}(k,\eta)\right|\e^{  \lambda(|k|+|\eta|)}<
+\infty$ with $\lambda>0$. If $\mu$ is small enough, then it exists a solution $f$ of \eqref{noneq} such that
\begin{align}
\sup_{\vartheta,\omega}\left|  f (t,\vartheta,\omega) - f _{\infty}(\vartheta-\omega t,\omega)\right|\xrightarrow[t\longrightarrow +\infty]{}0.
\end{align}
The convergence is  exponentially fast, and
$R(t)\leq C\e^{-\lambda t}$.
\end{theorem}
This theorem can be read as an existence result for the
kinetic Kuramoto equation with a prescribed asymptotic behavior.
A formulation of \eqref{noneq} in term of asymptotic data can be 
given by the use of the characteristics $\Theta(t,\vartheta,\omega)$:
\begin{equation}\label{weak}
(P)\,\, \left\{ 
\begin{aligned}
&\dot{\Theta}(t,\vartheta,\omega)=\omega-\mu R(t)\sin({\Theta}(t,\vartheta,\omega)-\varphi(t)),\,\, 
\\&\Theta(t,\vartheta,\omega)-\omega t \xrightarrow[t\to +\infty]{} \vartheta
\\
&R(t)\e^{\imm\varphi(t)}=
\int_{\Cal T\times \R} \e^{\imm \Theta(t,\vartheta,\omega)}f_\infty(\vartheta,\omega)d\vartheta d\omega \\
& f (t,\Theta(t,\vartheta, \omega),\omega)= f _\infty(\vartheta, \omega)\e^{-\mu \int_t^\infty R(s)\cos(\Theta(s,\vartheta,\omega)-\varphi(s)) ds}\
\end{aligned}\right..
\end{equation}
We want to construct a solution of the problem $(P)$ by doing the limit
of the solutions of a sequence of the linear problems. Starting with
$R_0=0$, we shall solve iteratively the following systems:
\begin{equation*}\label{weakn}
(P_n)\!\left\{ 
\begin{aligned}
&\dot{\Theta}_{n}(t,\vartheta,\omega)\!=\!\omega\!-\!\mu R_{n-1}(t)\sin({\Theta}_{n}(t,\vartheta,\omega)\!-\!\varphi_{n-1}(t)),
\\ 
&\Theta_{n}(t,\vartheta,\omega)\!-\!\omega t \!\xrightarrow[t\to +\infty]{} \!\vartheta
\\
&R_{n}(t)\e^{\imm\varphi_{n}(t)}=
\int_{\Cal T\times \R} \e^{\imm\Theta_n(t,\vartheta,\omega)}f_\infty(\vartheta,\omega)d\vartheta d\omega \\
& f (t,\Theta_n(t,\vartheta, \omega),\omega)= f _\infty(\vartheta, \omega)\e^{-\mu \int_t^\infty R_{n-1}(s)\cos(\Theta_n(s,\vartheta,\omega)-\varphi_{n-1}(s)) ds}\
\end{aligned}\right.\!.
\end{equation*}
In order to prove the existence and the 
convergence of the solution of $(P_n)$, we prove
some linear estimates in suitable functional spaces.
We define, for $\lambda > 0$:
\begin{align}
&\mathrm{X}_{\lambda}=\left\{ h:\R^+\times \Cal T\times \R\to \C \text{ s.t.} \sup_{t,\vartheta,\omega}\left|h(t,\vartheta,\omega)\e^{ \lambda t }\right| \leq C\right\};\\
&\left|\left|  h  \right|\right|_{\lambda}=\sup_{t,\vartheta,\omega}\left|h(t,\vartheta,\omega)\e^{\lambda  t }\right|. 
\end{align}
Given $R_{n-1}(t)\e^{\imm \varphi_{n-1}(t)}\in \mathrm{X}_{\lambda}$, we prove an existence and uniqueness result for the n-th characteristic:
\begin{align}\label{asymp}
\begin{cases}
\dot{\Theta}_{n}(t,\vartheta,\omega) =\omega -\mu R_{n-1}(t)\sin(\Theta_{n}(t,\vartheta,\omega)-\varphi_{n-1}(t))\\
\Theta_{n}(t,\vartheta,\omega)-\omega t \xrightarrow[t\to +\infty]{} \vartheta
\end{cases},
\end{align}
If a solution of \eqref{asymp} exists, it satisfies the  equation:
\begin{equation}
\label{eq:thetan}
\Theta_n(t,\vartheta,\omega)=\vartheta + \omega t + \mu \int_t^\infty R_{n-1}(s)\sin (\Theta_n(s,\vartheta,\omega)-\varphi_{n-1}(s))ds.
\end{equation}
We prove the existence by a contraction argument,  studying the 
following functional:
\begin{equation}
\Cal F_n(\Theta)=\vartheta + \omega t + \mu \int_t^\infty R_{n-1}(s)\sin (\Theta(s,\vartheta,\omega)-\varphi_{n-1}(s))ds.
\end{equation}
We shall use another functional space tailored on $\Cal F_n$: 
\begin{equation}
\tilde{\mathrm{X}}_{\lambda}=\left\{ \tilde h:\R^+\times \Cal T\times \R\to \C \text{ s.t. } \tilde h(t,\vartheta,\omega)-\vartheta -\omega t \in \mathrm{X}_{\lambda}  \right\}.
\end{equation}
\begin{proposition}
If $R_{n-1}\e^{\imm \varphi_{n-1}}\in {\mathrm{X}}_{\lambda}$ and $\mu$ is small enough, $\Cal F_n : \tilde{\mathrm{X}}_{\lambda}\longrightarrow \tilde{\mathrm{X}}_{\lambda}$ is a contractive operator.
\end{proposition}

\proof
$\Cal F_n$ is well defined on $\tilde{\mathrm{X}}_{\lambda}$:
\begin{align}
\left |\Cal F_n(\Theta) - \vartheta -\omega t\right|&\leq \mu \int_t^\infty R_{n-1}(s)ds \leq \mu  ||R_{n-1}||_{\lambda} \int_t^\infty \e^{-\lambda  s  }ds\leq\\
&\leq \frac{\mu}{\lambda} ||R_{n-1}||_{\lambda}\e^{-\lambda  t  }\nonumber;
\end{align}
multiplying by $\e^{\lambda t}$  and taking the $\sup$ on $t>0$ leads to 
\begin{align}
\left|\left |\Cal F_n(\Theta) - \vartheta -\omega t\right|\right|_{ \lambda }\leq  \frac{\mu}{\lambda}||R_{n-1}||_{\lambda},
\end{align}
which proves $\Cal F_n\left(\mathrm{X}_\lambda\right)\subseteq \mathrm{X}_\lambda$.

The operator $\Cal F_n$ is a contraction 
if $\frac{\mu}{\lambda}||R_{n-1} ||_\lambda<1$:
given $\Theta_1, \Theta_2$ in $\tilde{\mathrm{X}}_{\lambda}$
\begin{align}
\left |\Cal F_n(\Theta_1) -\Cal F_n(\Theta_2)\right|\leq&\Bigg|   
\mu \int_t^\infty \!\!\!\!\!\!\!R_{n-1}(s)\sin 
(\Theta_1(s,\vartheta,\omega)-\varphi_{n-1}(s))ds +\\
&\quad\quad- \mu \int_t^\infty \!\!\!\!\!\!\!R_{n-1}(s)\sin 
(\Theta_2(s,\vartheta,\omega)-\varphi_{n-1}(s))ds   \Bigg|\leq\nonumber\\
\leq& {\mu} \int_t^\infty \!\!\!\!\!\!\! R_{n-1}(s) | 
\Theta_1(s,\vartheta,\omega)\!-\! \Theta_2(s,\vartheta,\omega)| ds  
\leq\nonumber\\
\leq&\frac{\mu}{\lambda}  || \Theta_1\! -\!\Theta_2 ||_{\lambda} 
|| R_{n-1}||_{\lambda }{\e^{-  \lambda t }}.\nonumber
\end{align}
The thesis follows by multiplying by $\e^{\lambda t}$ and taking the 
$\sup$ on $t>0$.
\endproof
As a corollary, the system \eqref{asymp} has a unique solution $\Theta_n$ in $\tilde{\mathrm{X}}_{\lambda}$. Moreover, being  $\Theta_n$ 
the only fixed point of $\Cal F_n$, it satisfies
\begin{align}\label{estimrn}
\left|\left| \Theta_n -\vartheta -\omega t  \right|\right|_\lambda \leq \frac{\mu}{\lambda}||R_{n-1} ||_\lambda.
\end{align}

By the next lemma we show that if $R_{n-1}\e^{\imm \varphi_{n-1}}\in \mathrm{X}_{\lambda}$, then also $R_{n}\e^{\imm \varphi_{n}}\in\mathrm{X}_{\lambda}$.

\begin{lemma}\label{rbounde}
If $R_{n-1}\e^{\imm \varphi_{n-1}}\in \mathrm{X}_\lambda$ and 
$\sup_{k,\eta}\left|{\hat f}_{\infty}(k,\eta)\right|
\e^{  \lambda(|k|+|\eta|)}<+\infty$, the following estimate holds true:
\begin {equation}
|| R_{n} ||_{\lambda }\leq C\frac{\mu}{\lambda}\left[  ||R_{n-1} ||_{\lambda  }   + \sup_{k,\eta}\left|{\hat f}_{\infty}(k,\eta)\right|
\e^{\lambda(|k|+|\eta|)}\right].
\end{equation}
\end{lemma}

\proof
By the definition of $R_n\e^{\imm \varphi_n}$, it follows
\begin{align}\label{rfun}
R_n(t)\e^{\imm \varphi_n(t)}=&\int_{\Cal T\times \R} \e^{\imm \Theta_n(t,\vartheta,\omega) }f_\infty(\vartheta,\omega)d\vartheta d\omega=\\
=&\int_{\Cal T\times \R}\left[ \e^{\imm( \Theta_n(t,\vartheta,\omega) -\vartheta -\omega t)}-1\right]\e^{\imm(\vartheta +\omega t)}f _\infty(\vartheta,\omega)d\vartheta d\omega+\nonumber\\
&+\int_{\Cal T\times \R} \e^{\imm (\vartheta+\omega t) }f _\infty(\vartheta,\omega)d\vartheta d\omega.\nonumber
\end{align}
If $\alpha \in \R$, by $|\e^{\imm \alpha}-1|\leq |\alpha|$ 
we do a first estimate:
\begin{align}\label{da}
&\left|\int_{\Cal T\times \R}\left[ \e^{\imm (\Theta_n(t,\vartheta,\omega) -\vartheta -\omega t)}-1\right]\e^{\imm(\vartheta +\omega t)}f_\infty(\vartheta,\omega)d\vartheta d\omega\right|\leq \\
&\quad\leq \int_{\Cal T\times \R}\left|{ \Theta_n(t,\vartheta,\omega) -\vartheta -\omega t}\right| f _\infty(\vartheta,\omega)d\vartheta d\omega\leq\nonumber\\
&\quad\leq \frac{\mu}{\lambda} || R_{n-1}||_{ \lambda  }\e^{-\lambda  t  },
\nonumber
\end{align}
where the last inequality uses the estimate \eqref{estimrn}.
The second term of the r.h.s. in \eqref{rfun} is easily estimated: 
\begin{align}\label{db}
\left|\int _{\Cal T\times \R}\e^{\imm (\vartheta+\omega t) }
f_\infty(\vartheta,\omega)d\vartheta d\omega\right|=
\left| {\hat f}_\infty(-1,-t)
\right|\leq \e^{-\lambda  t}\sup_{k,\eta}\left|{\hat f}_\infty(k,\eta)
\right|\e^{  \lambda(|k|+|\eta|)}
\end{align}
The thesis is given by \eqref{rfun}, \eqref{da}, \eqref{db}.
\endproof

We proceed with the estimate of the difference between two consecutive characteristics.
\begin{lemma}\label{lemmin}
If $R_{n-1}\e^{\imm \varphi_{n-1}}$,  $R_{n-2}\e^{\imm \varphi_{n-2}}\in \mathrm{X}_\lambda$ and $\mu$ is sufficiently small, the following estimate holds true:
\begin{align}
||\Theta_{n} -\Theta_{n-1} ||_{\lambda }\leq \mu \frac{ ||R_{n-1}\e^{\imm \varphi_{n-1}}-R_{n-2}\e^{\imm \varphi_{n-2}}||_{\lambda }}{\lambda \left[1-\frac{\mu}{\lambda}||R_{n-1}||_{\lambda } \right]}.
\end{align}
\end{lemma}

\proof
Recalling the definition of $\Theta_n$ as solution of the equation
\ref{eq:thetan}:
\begin{align}\label{precede}
&\left|{ \Theta_{n}(t,\vartheta,\omega) } - { \Theta_{n-1}(t,\vartheta,\omega) }\right|\leq\\
&\leq \mu \int_t^\infty \!\!\!\Big|R_{n-1}(s)\sin(\Theta_n(s,\vartheta,\omega)-\varphi_{n-1}(s))+\nonumber\\
&\quad\quad\quad\quad\quad-R_{n-2}(s)\sin(\Theta_{n-1}(s,\vartheta,\omega)-\varphi_{n-2}(s))\Big|ds\leq\nonumber\\
&\leq \mu \int_t^\infty \!\!\!\left|R_{n-1}(s)\e^{\imm(\Theta_n(s,\vartheta,\omega)-\varphi_{n-1}(s))}-R_{n-2}(s)\e^{\imm(\Theta_{n-1}(s,\vartheta,\omega)-\varphi_{n-2}(s))}\right|ds.\nonumber
\end{align}
Summing and subtracting $R_{n-1}\e^{+\imm \Theta_{n-1}-\imm \varphi_{n-1}}$ in the absolute value and using the triangular inequality, we 
have
\begin{align}
| \Theta_{n}(t,\vartheta,&\omega)  - { \Theta_{n-1}(t,\vartheta,\omega) }|\leq\\
\leq &\mu \int_t^\infty \!\!\!\!\! \left|\e^{\imm \Theta_{n-1}(s,\vartheta,\omega)}\right| \left|R_{n-1}(s)\e^{-\imm \varphi_{n-1}(s)}-R_{n-2}(s)\e^{-\imm \varphi_{n-2}(s)}\right|ds+\nonumber\\
&+\mu \int_t^\infty \!\!\!\!\!\!\left|R_{n-1}(s)\e^{\imm \varphi_{n-1}(s)}\right| \left|\e^{\imm \Theta_{n}(s,\vartheta,\omega)}-\e^{\imm \Theta_{n-1}(s,\vartheta,\omega)} \right|ds\leq \nonumber\\
\leq& \frac{\mu}{\lambda} \left|\left|R_{n-1}(s)\e^{\imm \varphi_{n-1}(s)} -R_{n-2}(s)\e^{\imm \varphi_{n-2}(s)} \right|\right|_{\lambda }\e^{-\lambda t}+\nonumber\\
&+\frac{\mu}{\lambda}||R_{n-1}||_{ \lambda }||\Theta_{n} -\Theta_{n-1} ||_{\lambda }\e^{-\lambda t}\nonumber;
\end{align}
that proves to the thesis.
\endproof

By this last lemma we can 
estimate the difference of two consecutive $R_j\e^{\imm \varphi_j}$.

\begin{lemma}\label{cauc}
If $R_{n-1}\e^{\imm \varphi_{n-1}}$,  $R_{n-2}\e^{\imm \varphi_{n-2}}\in \mathrm{X}_\lambda$ and $\mu$ is sufficiently small, the following estimate holds true:
\begin{align}
\left|\left| R_{n}\e^{\imm \varphi_{n}}- R_{n-1}\e^{\imm \varphi_{n-1}} \right|\right|_\lambda\leq C\mu  \frac{ ||R_{n-1}\e^{\imm \varphi_{n-1}}-R_{n-2}\e^{\imm \varphi_{n-2}}||_{ \lambda }}{\lambda \left[1-\frac{\mu }{\lambda}||R_{n-1}||_{\lambda } \right]}.\
\end{align}
\end{lemma}

\proof

By Lemma \ref{lemmin}
\begin{align}
&\left| R_{n}(t)\e^{\imm \varphi_{n}(t)}- R_{n-1}(t)\e^{\imm \varphi_{n-1}(t)} \right|=\\
&\quad\quad\quad=\left| \int \left[\e^{\imm \Theta_{n}(t,\vartheta,\omega) } - \e^{\imm \Theta_{n-1}(t,\vartheta,\omega) }\right]f _\infty(\vartheta,\omega)d\vartheta d\omega \right|\leq\nonumber\\
&\quad\quad\quad\leq \int \left|{ \Theta_{n}(t,\vartheta,\omega) } - { \Theta_{n-1}(t,\vartheta,\omega) }\right| f _\infty(\vartheta,\omega)d\vartheta d\omega\leq\nonumber\\
&\quad\quad\quad\leq \mu  \frac{ ||R_{n-1}\e^{\imm \varphi_{n-1}}-R_{n-2}\e^{\imm \varphi_{n-2}}||_{ \lambda }}{\lambda \left[1-\frac{\mu }{\lambda}||R_{n-1}||_{\lambda } \right]}\e^{-\lambda t}\nonumber;
\end{align}
that  closes to the thesis.
\endproof

\noindent
We eventually give the proof of the main theorem of the section.

{\it Proof of Theorem \ref{backwardld}}.

\noindent
Given $f_\infty$ as in the hypothesis and $R_0=0$, for any $n\in \N$, $(P_n)$ 
has a unique solution $(R_n\e^{\imm \varphi_n}, \Theta_n, f_n)$.
First we prove by induction that $R_n$ is a bounded sequence in $\mathrm{X}_{\lambda}$.
$R_0$ is bounded by hypothesis, let $||R_m||_\gamma<M$ for 
$m\leq n-1$ and 
$$M>2\frac{C\mu}{\lambda} \sup_{k,\eta}\left[\left|\hat{f}_\infty(k,\eta)
\right|\e^{  \lambda(|k|+|\eta|)}\right],$$ 
with $C$ as in Lemma \ref{rbounde}.
In this case we have, by Lemma \ref{rbounde}: 
\begin{align}
||R_{n}||_\lambda&< M\left[\frac{1}{2}  +  \frac{C\mu}{\lambda}\right].
\end{align}
which, if $\mu$ is sufficiently small, proves that $||R_{n}||_\lambda<M$.\\
By the boundness of $R_n$, we have a lower bound of $\lambda \left[1-\frac{\mu }{\lambda}||R_{n-1}||_{\lambda } \right]$ for any $n\in \N$; again, taking $\mu$ sufficiently small, by Lemma \ref{cauc} it is true that
\begin{align}
\left|\left| R_{n}\e^{\imm \varphi_{n}}- R_{n-1}\e^{\imm \varphi_{n-1}} \right|\right|_{ \lambda } \leq \frac{1}{2} { ||R_{n-1}\e^{\imm \varphi_{n-1}}-R_{n-2}\e^{\imm \varphi_{n-2}}||_{ \lambda }},\
\end{align}
which proves that $R_n\e^{\imm \varphi_n}$ is a Cauchy sequence in $\mathrm{X}_{\lambda}$, then
\begin{align}
R_{n}(t)\e^{\imm \varphi_{n}(t)} \xrightarrow[n\longrightarrow 
+\infty]{\mathrm{X}_{\lambda}} R(t)\e^{\imm \varphi(t)}
\end{align}
and
\begin{align}
\Theta_n(t,\vartheta,\omega) \xrightarrow[n\longrightarrow 
+\infty]{\tilde{\mathrm{X}}_{\lambda}} \Theta(t,\vartheta,\omega).
\end{align}

Now we can define the solution $f$ of the problem $(P)$ as
\begin{align}
 f (t,\Theta(t,\vartheta, \omega),\omega)= f _\infty(\vartheta, \omega)\e^{-\mu \int_t^\infty R(s)\cos(\Theta(s,\vartheta,\omega)-\varphi(t)) ds};
\end{align}
clearly it is true that
$f_n \xrightarrow[n\longrightarrow +\infty]{} f$, as $n\to +\infty$.
We are left with proving that $f$ is asymptotically close to 
$f_\infty(\vartheta-\omega t, \omega)$.
\begin{align}
&\sup_{\vartheta,\omega}| f (t,\vartheta,\omega)- f _{\infty}(\vartheta-\omega t,\omega)|=\\
&= \sup_{\vartheta,\omega} | f (t,\Theta(t,\vartheta, \omega),\omega)- f _\infty(t,\Theta(t,\vartheta, \omega)-\omega t,\omega) |=\nonumber\\
&=| f _\infty(\vartheta, \omega)\e^{-\mu \int_t^\infty R(s)\cos(\Theta(s,\vartheta,\omega)-\varphi(t)) ds}- f _\infty(t,\Theta(t,\vartheta, \omega)-\omega t,\omega)|=\nonumber\\
&=|| f _\infty||_{L^\infty}|\e^{-\mu \int_t^\infty R(s)\cos(\Theta(s,\vartheta,\omega)-\varphi(t)) ds}-1|+\nonumber\\
&\quad+||\grad f _\infty||_{L^\infty}| |\Theta(s,\vartheta,\omega)-\vartheta-\omega t) |\leq\nonumber\\
&\leq|| f _\infty||_{L^\infty}C\mu \frac{\e^{C||R ||_{\lambda}}}{||R ||_{\lambda}+1}||R ||_{\lambda}\e^{-\lambda t}+\nonumber\\
&\quad+||\grad f _\infty||_{L^\infty}| ||\Theta(s,\vartheta,\omega)-\vartheta-\omega t) ||_{\lambda}\e^{-\lambda t},\nonumber
\end{align}
which vanishes to zero exponentially fast.
\endproof

\section{Polynomial dephasing}\label{sez:poly}

In this section we prove a second dephasing result where the
asymptotic datum $f_\infty$ is $\ve C^p$-regular and the decaying of
$R$ is only polynomial.  
Defining
$$\langle x\rangle=(1+x^2)^{\frac{1}{2}}, \quad \langle k,\eta \rangle=(1+|k|^2+|\eta|^2)^{\frac{1}{2}},$$
the precise statement of the result is:

\begin{theorem}\label{backwardldp}
Let $f_\infty: \Cal T\times \R\longrightarrow \R^+$ be a 
probability density which satisfies \\
$\sup_{k,\eta}\left|\hat{f}_\infty(k.\eta)\right|\ 
\langle k,\eta \rangle^\gamma<+\infty$ with $\gamma\geq2$. If $\mu$ 
is small enough, then it exits a unique 
solution $f$ of \eqref{noneq} such that
\begin{align}
\sup_{\vartheta,\omega}\left|  f (t,\vartheta,\omega) - f _{\infty}(\vartheta-\omega t,\omega)\right|\xrightarrow[t\longrightarrow +\infty]{}0;
\end{align}
the convergence is polynomial, and 
$R(t)\leq C \langle t \rangle^{-\gamma}$.
\end{theorem}

The construction of the solution is done by iteration in suitable
Banach spaces, and the convergence is proved by showing that the sequence of the solutions
of the problems $(P_n)$ is of Cauchy type; this last statement comes
after some linear estimates that are the contents of  Lemmas
\ref{rbo}, \ref{rba}, \ref{rbi}.

We give some short-hand notation to make more readable the proof:
\begin{align}
&\Gamma_n(t,\vartheta,\omega)=\int_t^\infty R_{n-1}(s)\sin(\Theta_n(t,\vartheta,\omega)-\varphi_n(s))ds;\\
&\Delta \Gamma_n(t,\vartheta,\omega)=\Gamma_n(t,\vartheta,\omega)-\Gamma_{n-1}(t,\vartheta,\omega);\\
&z_n(t)=R_n(t)\e^{\imm \varphi_n(t)}; \\
&\Delta Z_n(t)=R_n(t)\e^{\imm \varphi_n(t)}-R_{n-1}(t)\e^{\imm \varphi_{n-1}(t)};\\
&\beta_n(t)=\int_t^\infty R_n(s)ds.\label{beti}
\end{align}

The fixed point equation that arises from the problem
 $(P_n)$ can be rewritten as
\begin{align}
\Theta_n(t,\vartheta,\omega)=\Cal F_n(\Theta_n)=\vartheta+\omega t+ \Gamma_n(t,\vartheta,\omega);
\end{align}
by the definition \eqref{beti} it follows a simple estimate for $\Gamma_n$:
\begin{align}
\left|\Gamma_n(s,\vartheta,\omega)\right|\leq \beta_{n-1}(s)\label{disbeta}.
\end{align}
We define the functional space where we shall perform our iterative procedure
\begin{align}
\mathrm{Y}_\gamma=\left\{ h: \R^+\times\Cal T\times \R\longrightarrow \C \text{ s.t. } \sup_{t,\vartheta,\omega} \left|h(t,\vartheta,\omega)\right| \langle t\rangle^\gamma <+\infty  \right\},
\end{align}
\begin{align}\label{normagamma}
|| h||_\gamma=\sup_{t,\vartheta,\omega}|h(t,\vartheta,\omega)| \langle t\rangle^\gamma.
\end{align}
As in the previous section we define
\begin{align}
\tilde{\mathrm{Y}}_\gamma=\left\{ \tilde{h}: \R^+\times\Cal T\times \R\longrightarrow \C \text{ s.t. }  \tilde{h}(t,\vartheta,\omega)-\vartheta-\omega t \in \mathrm{Y}_\gamma  \right\}.
\end{align}
To start our iterative program we need these results 
for the functional $\Cal F_n$.
\begin{proposition}\label{exist}
If $R_{n-1}\e^{\imm \varphi_{n-1}}\in {\mathrm{Y}}_{\gamma}$ and $\mu$ is small enough, $\Cal F_n : \tilde{\mathrm{Y}}_{\gamma-1}\longrightarrow \tilde{\mathrm{Y}}_{\gamma-1}$ is a contractive operator.
\end{proposition}
\begin{corollary}
Given $R_{n-1}\e^{\imm \varphi_{n-1}}\in {\mathrm{Y}}_{\gamma}$ 
and $\mu$ sufficiently small, the only fixed point  $\Theta_n$ of $\Cal F_n$ is the unique solution of the problem $(P_n)$. Moreover $\Theta_n$ satisfies the following estimate:
\begin{align}
||\Theta_n -\vartheta-\omega t ||_{\gamma-1}\leq C {\mu} ||R_{n-1} ||_{\gamma}.
\end{align}
\end{corollary}
The corollary is a direct consequence of Prop. \ref{exist}. 

{\it Proof of Proposition \ref{exist}.}
\noindent
$\Cal F_n $ is well defined on $\tilde{\mathrm{Y}}_{\gamma-1}$
\begin{align}
|\Cal F_n(\Theta)-\vartheta-\omega t|\leq& \mu \,\beta_{n-1}(t)=\\
=&\mu\int_t^\infty R_{n-1}(s)ds \leq \mu  ||R_{n-1} ||_{\gamma}\int_{t}^\infty \frac{1}{ \langle s\rangle^{\gamma} }ds\leq\nonumber \\
\leq& C\mu  ||R_{n-1} ||_{\gamma}{ \langle t\rangle^{-\gamma+1} }.\nonumber
\end{align}
If $\frac{C\mu}{\lambda}||R_{n-1} ||_\gamma<1$, 
$\Cal F_n$ is contractive, as follows multiplying by 
$\langle t \rangle^{\gamma -1}$ the following inequality:
\begin{align}
\left |\Cal F(\Theta_1) -\Cal F(\Theta_2)\right|\leq&\Bigg|   \mu \int_t^\infty \!\!\!\!\!\!\!R_{n-1}(s)\sin (\Theta_1(s,\vartheta,\omega)-\varphi_{n-1}(s))ds  +\\
&\quad- \mu \int_t^\infty \!\!\!\!\!\!\!R_{n-1}(s)\sin (\Theta_2(s,\vartheta,\omega)-\varphi_{n-1}(s))ds   \Bigg|\leq\nonumber\\
\leq& {\mu} \left|\int_t^\infty \!\!\!\!\!\!\! R_{n-1}(s) | \Theta_1(s,\vartheta,\omega)\!-\! \Theta_2(s,\vartheta,\omega)| ds  \right|\leq\nonumber\\
\leq&\frac{C\mu}{\lambda}  || \Theta_1\! -\!\Theta_2 ||_{\gamma-1} || R_{n-1}||_{\gamma }{ \langle t\rangle }^{-\gamma+1}.\nonumber
\end{align}
\endproof

In the next lemmas, we shall often use the first and the
second order approximation of the phases:
\begin{align}
&1^{st}\,\,\text{order}\quad\e^{\imm x}=1+\e^{\imm \xi}x,\quad \xi \in \R;
\nonumber \\
&2^{nd}\,\,\text{order}\quad\e^{\imm x}=1+\imm x -\e^{\imm \xi}\frac{x^2}{2},\quad \xi \in \R.
\nonumber \end{align}
where $\xi$ is real and then $|\e^{\imm \xi}|\le 1$.

Now we present an estimate for $R_n(t) \e^{\imm \varphi_n(t)}$ in the norm \eqref{normagamma}.
\begin{lemma}\label{rbo}
Given $R_{n-1}\e^{\imm \varphi_{n-1}}\in {\mathrm{Y}}_{\gamma}$ and 
$\sup_{k,\eta}\left|
{\hat f}_\infty(k,\eta)\right| 
\langle \eta\rangle^\gamma <+\infty$ the following estimate holds true. 
\begin{align}
|| R_n||_\gamma \leq C\left[ 
\sup_{k,\eta}\left|
{\hat f}(k,\eta) \right|
\langle \eta\rangle^\gamma+\mu ||R_{n-1} ||_\gamma+\mu^2 ||R_{n-1} ||^2_\gamma\right]
\end{align}
\end{lemma}
\proof
By the first order approximation of the phase,  it is true that:
\begin{align}
R_n(t) \e^{\imm \varphi_n(t)} =&\int_{\Cal T\times \R} \e^{\imm \Theta_n(t,\vartheta,\omega) }f_\infty(\vartheta,\omega)d\vartheta d\omega=\\
=&\int_{\Cal T\times \R}  f_\infty(\vartheta,\omega)\e^{\imm (\vartheta+\omega t)}d\vartheta d\omega+\label{a0}\\
&+\imm \mu \int_{\Cal T\times \R}  f_\infty(\vartheta,\omega)\e^{\imm (\vartheta+\omega t)}\Gamma_n(t,\vartheta,\omega) d\vartheta d\omega+\label{a1} \\
&-\frac{\mu^2}{2} \int_{\Cal T\times \R}  f_\infty(\vartheta,\omega)\e^{\imm \xi}\Gamma_n(t,\vartheta,\omega)^2 d\vartheta d\omega\label{a2}
\end{align}
We start with an estimate for the term in \eqref{a0}
\begin{align}
\left|\int_{\Cal T \times \R}  f_\infty(\vartheta,\omega)\e^{\imm (\vartheta+\omega t)}d\vartheta d\omega\right|=\left| 
{\hat f}_\infty (-1,-t)\right| 
\leq \sup_{k,\eta}\left[\left| {\hat f}_{\infty}(k,\eta)\right| 
\langle \eta\rangle^\gamma \right]\langle t\rangle^{-\gamma},
\end{align}
then we bound the quantity in \eqref{a2}
\begin{align}
\left| -\frac{\mu^2}{2}\int_{\Cal T \times \R}  f_\infty(\vartheta,\omega)\e^{\imm \xi}\Gamma_n(t,\vartheta,\omega)^2 d\vartheta d\omega  \right|\leq \frac{\mu^2}{2} \beta_{n-1}^2=\frac{\mu^2}{2}\left(  \int_t^\infty R_{n-1}(s)ds  \right)^2\leq\\
\leq \frac{\mu^2}{2}|| R_{n-1}||^2_\gamma \left(\int_t^\infty \frac{1}{ \langle s\rangle^\gamma }ds\right)^2\leq C\mu^2 \langle t\rangle^{-2\gamma+2}|| R_{n-1}||^2_\gamma.\nonumber
\end{align}
We rewrite the term in \eqref{a1} using the definition of $\Gamma_n$ and the Euler's identity:
\begin{align}
\imm \mu \int_{\Cal T\times \R}  f_\infty(\vartheta,\omega)\e^{\imm (\vartheta+\omega t)}\Gamma_n(t,\vartheta,\omega) d\vartheta d\omega=\frac{\mu}{2}\int_{\Cal T \times \R} d\vartheta d\omega f_\infty(\vartheta,\omega)\e^{\imm (\vartheta+\omega t)}\cdot\\
\cdot \int_t^\infty dsR_{n-1}(s)\left[ \e^{\imm (\Theta_n(s,\vartheta,\omega) -\varphi_{n-1}(s) )} -\e^{-\imm (\Theta_n(s,\vartheta,\omega) -\varphi_{n-1}(s) )} \right],\nonumber
\end{align}
then we decompose the r.h.s of the last identity in the sum of two addends:
\begin{align}
&A_1^+=\frac{\mu}{2}\int_t^\infty ds R_{n-1}(s)\e^{-\imm \varphi_{n-1}(s)}\int_{\Cal T\times \R} d\vartheta d\omega f_\infty(\vartheta,\omega)\e^{\imm(2\vartheta +\omega(t+s))+\imm \mu \Gamma_n(s,\vartheta,\omega))},\\
&A_1^-=-\frac{\mu}{2}\int_t^\infty ds R_{n-1}(s)\e^{+\imm \varphi_{n-1}(s)}\int_{\Cal T\times \R} d\vartheta d\omega f_\infty(\vartheta,\omega)\e^{\imm(\omega(t-s)-\imm \mu \Gamma_n(s,\vartheta,\omega))}.
\end{align}
By the expansion of $\e^{\imm x}$ to the first order:
\begin{align}
A_1^+=&\underbrace{\frac{\mu}{2}\int_t^\infty ds R_{n-1}(s)\e^{-\imm \varphi_{n-1}(s)}\int_{\Cal T\times \R} d\vartheta d\omega f_\infty(\vartheta,\omega)\e^{\imm(2\vartheta +\omega(t+s))}}_{A^+_{1,0}}+\\
&+\underbrace{\imm \frac{\mu^2}{2}\int_t^\infty ds R_{n-1}(s)\e^{-\imm \varphi_{n-1}(s)}\int_{\Cal T\times \R} d\vartheta d\omega f_\infty(\vartheta,\omega)\e^{\imm \xi}\Gamma_{n}(s,\vartheta,\omega)}_{{A^+_{1,1}}},\nonumber
\end{align}
\begin{align}
A_1^-=&\underbrace{\frac{\mu}{2}\int_t^\infty ds R_{n-1}(s)\e^{\imm \varphi_{n-1}(s)}\int_{\Cal T\times \R} d\vartheta d\omega f_\infty(\vartheta,\omega)\e^{\imm\omega(t-s)}}_{A^-_{1,0}}+\\
&\underbrace{-\imm \frac{\mu^2}{2}\int_t^\infty ds R_{n-1}(s)\e^{\imm \varphi_{n-1}(s)}\int_{\Cal T\times \R} d\vartheta d\omega f_\infty(\vartheta,\omega)\e^{\imm \xi}\Gamma_{n}(s,\vartheta,\omega)}_{A^-_{1,1}}.\nonumber
\end{align}
The estimate for $A_{1,1}^\pm$ is easier:
\begin{align}
\left| A_{1,1}^\pm  \right|\leq \frac{\mu^2}{2}\beta_{n-1}(t)^2\leq C\mu^2 \langle t\rangle^{-2\gamma+2}|| R_{n-1}||_\gamma^2.
\end{align}
Using the definition of the Fourier transform it is clear that:
\begin{align}
&A_{1,0}^+ ={\mu}\pi\int_t^{\infty}ds R_{n-1}(s)\e^{-\imm\varphi_{n-1}(s)}
{\hat f}_\infty(-2,-t-s),\\
&A_{1,0}^- ={\mu}\pi\int_t^{\infty}ds R_{n-1}(s)\e^{\imm\varphi_{n-1}(s)}
{\hat f}_\infty(0,s-t).
\end{align}
By taking the absolute value in the previous identities, it follows:
\begin{align}
\left|  A^+_{1,0} \right|\leq &\pi\mu\sup_{k,\eta}\left|{\hat f}_{\infty}(k,\eta) \langle \eta\rangle^\gamma \right| \int_t^\infty ds R_{n-1}(s) \langle t+s\rangle^{-\gamma}\leq\\
\leq& \pi\mu \sup_{k,\eta}\left|{\hat f}_{\infty}(k,\eta) \langle \eta\rangle^\gamma \right|  ||R_{n-1} ||_\gamma \langle t\rangle^{-\gamma}\int_{t}^\infty \frac{ \langle t\rangle^\gamma  }{ \langle s\rangle^\gamma \langle t+s\rangle^{\gamma}  }ds\leq\nonumber\\
\leq& C\mu \sup_{k,\eta}\left|{\hat f}_{\infty}(k,\eta) \langle \eta\rangle^\gamma \right|   || R_{n-1} ||_{\gamma} \langle t\rangle^{-\gamma};\nonumber
\end{align}

\begin{align}
\left|  A^-_{1,0} \right|\leq&\pi\mu \sup_{k,\eta}\left|{\hat f}_{\infty}(k,\eta) \langle \eta\rangle^\gamma \right|  \int_t^\infty ds R_{n-1}(s) \langle s-t\rangle^{-\gamma}\leq\\
\leq& \pi\mu\sup_{k,\eta}\left|{\hat f}_{\infty}(k,\eta) \langle \eta\rangle^\gamma \right|   ||R_{n-1} ||_\gamma \langle t\rangle^{-\gamma}\int_{t}^\infty \frac{ \langle t\rangle^\gamma  }{ \langle s\rangle^\gamma \langle s-t\rangle^{\gamma}  }ds\leq\nonumber\\
\leq& C\mu \sup_{k,\eta}\left|{\hat f}_{\infty}(k,\eta) \langle \eta\rangle^\gamma \right|   || R_{n-1} ||_{\gamma} \langle t\rangle^{-\gamma}.\nonumber
\end{align}
The thesis  follows by the estimate we just proved. 
\endproof

\begin{lemma}\label{rba}
Let $R_{n-1}\e^{\imm \varphi_{n-1}},R_{n-2}\e^{\imm \varphi_{n-2}}\in {\mathrm{Y}}_\gamma$,  if $\mu$ is sufficiently small, it is true that

\begin{align}
||\Delta \Gamma_n ||_{\gamma-1}\leq\frac{C}{1-\mu C||R_{n-2}||_\gamma} ||\Delta Z_{n-1}||_{\gamma} 
\end{align}

\end{lemma}

\proof
Using the definition of $\Delta \Gamma_n$
\begin{align}
\Delta\Gamma_n(t,\vartheta,\omega)=\Gamma_{n}&(t,\vartheta,\omega)- \Gamma_{n-1}(t,\vartheta,\omega)=\\
=&\int_t^\infty ds R_{n-1}(s)\sin(\Theta_{n}(s,\vartheta,\omega)-\varphi_{n-1}(s,\vartheta,\omega))+\nonumber\\
&-\int_t^\infty dsR_{n-2}(s)\sin(\Theta_{n-1}(s,\vartheta,\omega)-\varphi_{n-2}(s,\vartheta,\omega))=\nonumber\\
=&\frac{1}{2\imm}\int_t^\infty ds\left[  R_{n-1}(s)\e^{-\imm \varphi_{n-1}(s)} - R_{n-2}(s)\e^{-\imm \varphi_{n-2}(s)} \right]\e^{\imm \Theta_n(s,\vartheta,\omega)}+\nonumber\\
&+ \frac{1}{2\imm}\int_t^\infty ds R_{n-2}(s)\e^{-\imm \varphi_{n-2}(s)}\left[  \e^{\imm \Theta_n(s,\vartheta,\omega)}-\e^{\imm \Theta_{n-1}(s,\vartheta,\omega)} \right]+\nonumber\\
&-\frac{1}{2\imm}\int_t^\infty ds\left[  R_{n-1}(s)\e^{\imm \varphi_{n-1}(s)} - R_{n-2}(s)\e^{\imm \varphi_{n-2}(s)} \right]\e^{-\imm \Theta_n(s,\vartheta,\omega)}+\nonumber\\
&-\frac{1}{2\imm}\int_t^\infty ds R_{n-2}(s)\e^{\imm \varphi_{n-2}(s)}\left[  \e^{-\imm \Theta_n(s,\vartheta,\omega)}-\e^{-\imm \Theta_{n-1}(s,\vartheta,\omega)} \right];\nonumber
\end{align}
then, taking the absolute value in the previous identity we have
\begin{align}
&|\Delta\Gamma_{n}(t,\vartheta,\omega))|\leq\\
&\int_t^\infty ds|z_{n-1}(s)-z_{n-2}(s)|+\mu \int_t^\infty ds|R_{n-2}(s)| |\Gamma_{n}(s,\vartheta,\omega)-\Gamma_{n-1}(s,\vartheta,\omega) |\leq\nonumber\\
&\leq C ||\Delta Z_{n-1}||_\gamma \langle t \rangle^{-\gamma+1} +C\mu ||R_{n-2}||_\gamma || \Delta \Gamma_n||_{\gamma-1}  \langle t \rangle^{-2\gamma+2}.\nonumber
\end{align}
Multiplying by $ \langle t \rangle^{\gamma-1}$ and taking the $\sup$ on $t$ 
gives
\begin{align}
||\Delta \Gamma_n ||_{\gamma-1}\leq C ||\Delta Z_{n-1}||_{\gamma}+\mu C ||R_{n-2}||_\gamma || \Delta \Gamma_n||_{\gamma-1},
\end{align}
which proves the thesis.
\endproof

We prove the last lemma that shows an estimate for $\Delta Z_n$.

\begin{lemma}\label{rbi}
Let $M>0$ such that $\sup_{k,\eta}\left|{\hat f}_{\infty}(k,\eta) \langle \eta\rangle^\gamma \right|\leq M$ and\\ $\sup_{j=1,\dots, n} ||R_j\e^{\imm \varphi_j}||_\gamma \leq M$; then, if $\mu$ is sufficiently small, the following estimate holds true

\begin{align}
||\Delta Z_n ||_\gamma\leq C M\left[\mu||\Delta Z_{n-1}||_\gamma+\mu^2 ||\Delta Z_{n-1} ||_\gamma\right]
\end{align}

\end{lemma}
\proof
The quantity $\Delta Z_n$ can be rewritten as follows:
\begin{align}
&\Delta Z_n(t)=\int_{\Cal T\times \R} d\vartheta d\omega f_\infty(\vartheta,\omega)\left[\e^{\imm \Theta_n(t,\vartheta,\omega)}-\e^{\imm \Theta_{n-1}(t,\vartheta,\omega)}\right]=\\
&\quad=\underbrace{\int_{\Cal T\times \R} d\vartheta d\omega f_\infty(\vartheta,\omega)\e^{\imm (\vartheta+\omega t)}\left( \e^{\imm \mu (\Gamma_{n}(t,\vartheta,\omega)-\Gamma_{n-1}(t,\vartheta,\omega))} -1 \right)}_{B_1}\nonumber+\\
&\quad+\underbrace{\int_{\Cal T\times \R} \!\!d\vartheta d\omega f_\infty(\vartheta,\omega)\e^{\imm (\vartheta+\omega t)}\left( \e^{\imm \mu \Gamma_{n-1}(t,\vartheta,\omega)} -1\right)\left( \e^{\imm \mu (\Gamma_{n}(t,\vartheta,\omega)-\Gamma_{n-1}(t,\vartheta,\omega))} -1 \right)}_{B_2};\nonumber
\end{align}
taking the absolute values on $B_2$
\begin{equation}
|B_2|\leq \mu^2 \int_{\Cal T\times \R} d\vartheta d\omega f_\infty (\vartheta, \omega)|\Gamma_{n-1}(t,\vartheta,\omega)| |\Gamma_{n}(t,\vartheta,\omega)-\Gamma_{n-1}(t,\vartheta,\omega) |,
\end{equation}
by Lemma \ref{rba}
\begin{align}
|B_2|\leq \frac{C\mu^2||\Delta Z_{n-1} ||_\gamma ||R_{n-1}||_\gamma \langle t\rangle^{-2\gamma+2}
}{1-\mu C||R_{n-2}||_\gamma}.
 \end{align}
Using the second order expansion of the phase
\begin{align}
&B_1=B_{1,1}+B_{1,2},\\
&B_{1,1}=\imm \mu \int_{\Cal T\times \R} d\vartheta d\omega f_\infty(\vartheta,\omega)\e^{\imm(\vartheta+\omega t)}\left(\Gamma_{n}(t,\vartheta,\omega)-\Gamma_{n-1}(t,\vartheta,\omega)\right),\\
&B_{1,2}=-\mu^2 \int_{\Cal T\times \R} d\vartheta d\omega f_\infty(\vartheta,\omega) \e^{\imm (\vartheta+\omega t)}\e^{\imm \xi}\left(\Gamma_{n}(t,\vartheta,\omega)-\Gamma_{n-1}(t,\vartheta,\omega)\right)^2.
\end{align}

The estimate for $B_{1,2}$ is done as follows: 
\begin{align}
|B_{1,2}|\leq&  \mu^2  |  \Gamma_{n}(t,\vartheta,\omega)-\Gamma_{n-1}(t,\vartheta,\omega)|^2\leq\\
\leq& \mu^2\left[|\Gamma_{n}(t,\vartheta,\omega)|+|\Gamma_{n-1}(t,\vartheta,\omega)|  \right]\left|\Gamma_{n}(t,\vartheta,\omega)-\Gamma_{n-1}(t,\vartheta,\omega) \right|\leq\nonumber\\
\leq&C\mu^2\left[ ||R_n ||_\gamma+||R_{n-2} ||_\gamma\right]||\Delta\Gamma_n ||_{\gamma-1} \langle t \rangle^{-2\gamma+2},\nonumber
\end{align}
again by Lemma \ref{rba}:
\begin{align}
|B_{1,2}|\leq\frac{C\mu^2\left[ ||R_n ||_\gamma+||R_{n-1} ||_\gamma\right]}{1-\mu C||R_{n-2}||_\gamma} ||\Delta Z_{n-1}||_{\gamma}  \langle t\rangle^{-2\gamma+2}-
\end{align}

We are left with $B_{1,1}=\tilde{B}_0^++\tilde{B}_1^++\tilde{B}_0^-+\tilde{B}_1^-$, where:
\begin{align}
\tilde{B}_0^+ =&\mu \int_{\Cal T\times \R} d\vartheta d\omega f_\infty(\vartheta, \omega) \e^{\imm (\vartheta+\omega t)}\int_t^\infty ds \overline{\Delta Z_{n-1}(s)}\e^{\imm \Theta_{n}(s,\vartheta,\omega)};\\
\tilde{B}_0^- =&-\mu \int_{\Cal T\times \R} d\vartheta d\omega f_\infty(\vartheta, \omega) \e^{\imm (\vartheta+\omega t)}\int_t^\infty ds \Delta Z_{n-1}(s)\e^{-\imm \Theta_{n}(s,\vartheta,\omega)};\\
\tilde{B}_1^\pm=&\pm\mu \int_{\Cal T\times \R} d\vartheta d\omega f_\infty(\vartheta, \omega)\e^{\imm(\vartheta+\omega t)}\cdot\\
&\quad\quad\cdot\int_t^\infty ds R_{n-2}(s)\e^{\mp\imm\varphi_{n-2}(s)}\Big[  \e^{\pm\imm \Theta_n(s)}-\e^{\pm \imm \Theta_{n-1}(s,\vartheta,\omega)}  \Big].\nonumber
\end{align}

The easiest estimate is for $\tilde{B}_1^\pm$,
\begin{align}
\left|\tilde{B}_1^\pm\right|\leq& \mu ||R_{n-2} ||_{\gamma} \int_t^\infty  \langle s\rangle^{-\gamma} |\Delta\Gamma_{n}(s,\vartheta,\omega)|\leq\\
\leq& \mu || R_{n-2} ||_{\gamma}||\Delta \Gamma_n ||_{\gamma-1}\int_{t}^\infty \frac{ds}{\langle s \rangle^{\gamma} \langle s \rangle^{\gamma-1}}\leq\nonumber\\
\leq& \frac{C\mu || R_{n-2}||_\gamma}{1-\mu C||R_{n-2}||_\gamma} ||\Delta Z_{n-1}||_{\gamma} \langle t\rangle^{-2\gamma+2}.\nonumber
\end{align}

The more involved estimate is for $\tilde{B}_0^-$:
\begin{align}
\tilde{B}_0^-= -\mu \int_t^\infty ds \int_{\Cal T\times \R} d\vartheta d\omega f_\infty (\vartheta,\omega) \e^{\imm \omega(t-s)-\imm \mu \Gamma_{n}(s,\vartheta,\omega)}{\Delta Z_{n-1}(s)},
\end{align}
expanding the non linear part of the phase up to the first order one 
write $\tilde{B}_0^-=\tilde{B}_{0,0}^-+\tilde{B}_{0,1}^-$, where
\begin{align}
&\tilde{B}_{0,0}^-= -\mu \int_t^\infty ds \int_{\Cal T\times \R} d\vartheta d\omega f_\infty (\vartheta,\omega) \e^{\imm \omega (t-s)}{\Delta Z_{n-1}(s)},\\
&\tilde{B}_{0,1}^-=-\imm \mu^2\int_t^\infty ds \int _{\Cal T\times \R}d\vartheta d\omega f_\infty (\vartheta,\omega) \e^{\imm \omega(t-s)+\imm\xi}\Gamma_{n}(s,\vartheta,\omega){\Delta Z_{n-1}(s)}.
\end{align}
As usual, we bound  $\tilde{B}_{0,0}^-$ and $\tilde{B}_{0,1}^-$ 
by taking the absolute value in the integral:
\begin{align}
|\tilde{B}_{0,0}^-| =& \left|\mu \int_t^\infty ds {\Delta Z_{n-1}(s)} \int _{\Cal T\times \R}d\vartheta d\omega f_\infty (\vartheta,\omega) \e^{\imm \omega (t-s)}\right|\leq\\
\leq&\mu \int_t^\infty ds\left| {\Delta Z_{n-1}(s)}  \right|\left| 
{\hat f}_\infty (0,s-t) \right|\leq\nonumber\\
\leq& \mu||\Delta Z_{n-1} ||_\gamma \sup_{k,\eta}\left|{\hat f}_{\infty}(k,\eta) \langle \eta\rangle^\gamma \right| \int _t^\infty \frac{ds}{\langle s\rangle^\gamma\langle s-t\rangle^\gamma}\leq\nonumber\\
\leq& C\mu||\Delta Z_{n-1} ||_\gamma \sup_{k,\eta}\left|{\hat f}_{\infty}(k,\eta) \langle \eta\rangle^\gamma \right|\langle t \rangle^{-\gamma};\nonumber
\end{align}

\begin{align}
\left| \tilde{B}_{0,1}^- \right|\leq& \mu^2 \int_t^\infty ds \int_{\Cal T\times \R} d\vartheta d\omega f_\infty(\vartheta,\omega)|\Delta Z_{n-1}(s)| |\Gamma_{n}(s,\vartheta,\omega)|\leq\\
\leq& \mu^2 ||\Delta Z_{n-1} ||_\gamma ||R_{n-1} ||_\gamma \int_t^\infty \frac{ds} {\langle s\rangle ^{\gamma} \langle s\rangle^{\gamma -1}}\leq \mu^2 C || \Delta Z_{n-1}||_\gamma ||R_{n-1} ||_\gamma \langle t\rangle^{-2\gamma+2}.\nonumber
\end{align}

We omit the simpler proof of the 
estimate for $\tilde{B}^+_0$, because it follows by the same calculations we just did for the estimate of $\tilde{B}^-_0$, with $t+s$ instead of $t-s$ in the formulas.  

The thesis of the lemma comes by collecting the estimates we did above.

\endproof

{\it Proof of Theorem \ref{backwardldp}.}

\noindent
The proof proceeds like that of Theorem \ref{backwardld},
as a consequence of Lemmas \ref{rbo}, \ref{rba}, \ref{rbi}. 

\endproof

\end{document}